\documentclass[prb,twocolumn]{revtex4-1} 

\usepackage[breaklinks,colorlinks,citecolor=blue]{hyperref}

\usepackage{amsfonts}
\usepackage{amssymb}

\newcommand{\be}{\begin{equation}}
\newcommand{\ee}{\end{equation}}
\newcommand{\bea}{\begin{eqnarray}}
\newcommand{\eea}{\end{eqnarray}}

\begin{document}

\title[Energy conservation in explicit solutions as a simple illustration of Noether's theorem]{Energy conservation in explicit solutions as a simple illustration of Noether's theorem}

\author{Markus P{\"o}ssel}
\address{Haus der Astronomie and Max Planck Institute for Astronomy, K{\"o}nigstuhl 17, 69124 Heidelberg, Germany}
\email{poessel@hda-hd.de}

\begin{abstract}
Noether's theorem is widely regarded as one of the most elegant results in theoretical physics. The article presents two simple examples that can be used to demonstrate the basic idea behind Noether's theorem, by deriving a relation between translational symmetry in the time coordinate and energy conservation in a way that is accessible to beginning students.\\[1em]
Note: The following article has been accepted for publication in the {\em American Journal of Physics}. After it is published, it will be found at \href{https://aapt.scitation.org/journal/ajp}{https://aapt.scitation.org/journal/ajp} 
\end{abstract}

\maketitle 

\section{Introduction}
Noether's theorem, published by the mathematician Amalie (Emmy) Noether in 1918,\cite{Noether1918a,Noether1918b,Byers1996,Kosmann-Schwarzbach2011} is widely regarded as one of the most elegant, beautiful and powerful results in theoretical physics.\cite{Griffiths2008,LedermanHill2004,Wilczek2015} The theorem proves a deep relationship between symmetries and conserved quantities.\cite{DualNote}

In everyday life, a system possesses a symmetry whenever there are operations we can perform on the system that leave the system's essential properties unchanged. If, for example, we take an axisymmetric vase and rotate it by a fixed angle around its axis of symmetry, then after the operation the vase will look just as it looked before. This everyday notion of symmetry is very similar to the more technical definition used in physics: There, systems and their components are described in terms of coordinates. We can actively change a system's coordinates, which can in turn change key physical properties of the system. For instance, re-locating a particle within an external potential will, in general, change the particle's potential energy. An active coordinate transformation is called a symmetry if it does {\em not} change the system's physical properties, more specifically: leaves the Lagrangian that encodes the system's dynamics unchanged.

As for conserved quantities, students encounter examples in their introductory physics class in the shape of energy, linear and angular momentum.

The relation between symmetries and conserved quantities found by Noether is much more difficult to understand than the two basic concepts it connects. Proofs of Noether's theorem make use of the Lagrangian and Hamiltonian formalisms of classical mechanics and field theory, and require knowledge of advanced concepts such as functionals and stationarity. While these can be made accessible to advanced undergraduate students,\cite{Neuenschwander2017} and have found their way into undergraduate-level teaching in various interesting roles,\cite{Neuenschwander2014} the effort required for introducing the necessary tools puts such general treatments beyond the scope of beginning physics students, and beyond introductory courses that are part of a general science requirement.

In the following, I present two examples that can serve to illustrate a basic relation between symmetries and conserved quantities, analogous to the relation described by Noether's theorem, in a way that is accessible to beginning physics students, general science students and even advanced high school students. The calculations needed for the derivation require no more than working knowledge of one-dimensional Newtonian mechanics, basic algebra and trigonometry. 

The examples demonstrate a direct relation between energy conservation and constant shifts in the time coordinate. They exploit the fact that in explicit solutions of equations of motion, the time coordinate is the basic independent variable, so that one can implement shifts in this variable directly (not via an infinitesimal generator, but as transformations of explicit solutions). Explicit solutions can be written using no more than the time variable and suitable parameters that specify the initial conditions, and ``shape invariance'' of these solutions then leads directly to the usual expressions for energy conservation.

\section{Energy conservation for a test particle in a constant, homogeneous field}
Consider the motion of a test particle in a static, homogeneous field $f$ in one spatial dimension x. The equation of motion for a test particle with mass $m$ and field-specific charge $q$ in such a field is
\be
m\,\ddot{x} = q\,f,
\ee
and is readily integrated to yield
\be
\label{ConstantHomogeneousSolution}
x(t) = x_0 + v_0t + \frac12 a t^2,
\ee
where the constant acceleration $a$ is given by 
\be
a = \frac{q}{m}\cdot f.
\label{accelerationField}
\ee
The two integration constants parametrize the initial conditions, namely the location $x_0$ and the velocity in x-direction $v_0$ at time $t=0$. Let us call $x_0$ and $v_0$ the initial parameters of our system.

There is no explicit time dependence in our set-up. Thus, when we shift the time coordinate as $t\mapsto t' = t+\Delta t$, with some constant shift $\Delta t$, describing the exact same situation as in (\ref{ConstantHomogeneousSolution}), we must again be able to write the result in the form (\ref{ConstantHomogeneousSolution}), albeit in general with different values for the initial parameters, which we shall call $x_1$ and $v_1$:
\be
\label{ShiftedEquation}
x'(t') = x_1 + v_1t' + \frac12 a t'{}^2.
\ee
The prime on the $x$ indicates that this is now a different coordinate function than before. In this case, $x_1$ and $v_1$ are the (initial) location and (initial) speed at $t'=0$, that is, at a point in time distinct from $t=0$.
This ``shape invariance'' of the explicit solution is our simplified counterpart to the invariance of the Lagrangian in the usual definition of Noether's theorem. Substituting $t'=t+\Delta t$ into (\ref{ShiftedEquation}), we must recover the original solution (\ref{ConstantHomogeneousSolution}). In particular, after the substitution, (\ref{ShiftedEquation}) becomes an equation with a part that is constant in $t$, a part linear in $t$ and a part quadratic in $t$. For that equation to be equal to (\ref{ConstantHomogeneousSolution}) for all values of $t$, the constant, linear and quadratic parts must be equal, separately. This requires that the initial parameter values $v_1$ and $x_1$ must be related to the original initial parameter values as 
\bea
\label{vRelation}
v_1 &=& v_0 -a\,\Delta t\\
\label{xRelation}
x_1 &=& x_0 - v_0\Delta t + \frac12 a\Delta t^2.
\eea
These two equations can be combined to eliminate $\Delta t$, since, from (\ref{vRelation}), we have
\be
\label{DeltaTExpressed}
\Delta t = \frac{1}{a}(v_0-v_1).
\ee
Substitution of this result into (\ref{xRelation}) yields a relation between the initial parameters at the two different times $t'=0$ and $t=0$, namely
\be
x_0 = x_1 + \frac{1}{2a}(v_0^2-v_1^2).
\ee
By substituting the expression (\ref{accelerationField}) for the acceleration $a$, this can be re-written as 
\be
\frac12 m v_0^2 - qfx_0 = \frac12 m v_1^2 - qfx_1.
\ee
This equation does not depend explicitly on the shift $\Delta t$, and thus holds for positions $x_0, x_1$ and speeds $v_0,v_1$ at any two different points in time. Since we could have chosen any two times $t'=0$ and $t=0$ for our analysis, the relation must hold for locations and speeds evaluated at {\em any} specific moment. The terms on each side of the equation are the usual expression for the kinetic and potential energy of a particle in a static, homogeneous field. In other words: From the time-translation invariance of our general solution with arbitrary initial parameter values, we have derived the usual expression for energy conservation! 

The simplicity of the set-up, allowing for an explicit general solution in the first place, has led us to a special case of an integral version of Noether's theorem -- certainly not powerful enough to assist in any serious calculations, but simple enough to demonstrate the connection between symmetry and conservation laws to beginning students.

\section{Energy conservation for the harmonic oscillator}

As another example, consider the harmonic oscillator --- an ubiquitous model in physics, since it follows from the lowest-order approximation for the dynamics around a stable equilibrium state. In this case, the dynamical equation is linear in $x$ (Hooke's law):
\be
m\ddot{x} = -k\cdot x,
\ee
with $k>0$ the spring constant. We introduce the angular frequency $\omega$ by defining 
\be
\omega^2 = \frac{k}{m}.
\ee
The most general solution can be written as
\be
x(t) = x_0\cos(\omega t) + \frac{v_0}{\omega}\sin(\omega t).
\label{OscillatorSolution}
\ee
This explicitly shows the role of the initial parameters $x_0$ and $v_0$, which are again the initial position and initial speed. Our general argument remains the same: By time-translation invariance, any explicit solution with initial conditions specified at a different moment in time must have the same shape as 
 (\ref{OscillatorSolution}); expressed in the shifted time coordinate $t' = t+\Delta t$ and with the same notation as in the previous section,
\be
x'(t') = x_1\cos(\omega t') + \frac{v_1}{\omega}\sin(\omega t').
\label{OscillatorSolution2}
\ee
Once more, we can substitute $t' = t+\Delta t$, and shape invariance will mean that the result must have the same functional dependence on $t$ as the original (\ref{OscillatorSolution}). Just as in our earlier calculation, that fixes specific relations between the two sets of initial parameters. To see this, re-write $x'(t+\Delta t)$ as the sum of a term proportional to $\cos(\omega t)$ and a term proportional to $\sin(\omega t)$. This can be achieved by using the addition formulae for sine and cosine on the expression $x'(t+\Delta t)$. The requirement that the coefficients of the sine and cosine term be the same as in the original (\ref{OscillatorSolution}) leads to the two equations
\bea
 \label{HOx1}
x_1 &=& x_0\;\cos(\omega\,\Delta t) - \frac{v_0}{\omega}\;\sin(\omega\,\Delta t) \\
v_1 &=& \omega x_0\;\sin(\omega\,\Delta t) + v_0\;\cos(\omega\,\Delta t). \label{HOv1}
\eea
This is equivalent to an equation for $\Delta t$
\begin{widetext}
\be
\Delta t = \frac{1}{\omega} \arctan\left[
\frac{\omega(v_1x_0-v_0x_1)}{x_0x_1\omega^2+v_1v_0} \right]
= \frac{1}{\omega}\left[ \arctan\left(\frac{v_1}{\omega x_1}\right) -
\arctan\left(\frac{v_0}{\omega x_0}\right)\right], \label{DeltaTHarmonicOscillator}
\ee
\end{widetext}
on the one hand, and the $\Delta t$-independent equation
\begin{equation}
\omega^2x_0^2 + v_0^2 = \omega^2x_1^2 + v_1^2
\label{EnergyHO}
\end{equation}
on the other. The latter equation is obtained by calculating $x_1^2 + (v_1/\omega)^2$ using (\ref{HOx1}) and (\ref{HOv1}), and applying the simple trigonometric identity $\sin^2 \phi + \cos^2 \phi=1$. When multiplied with the constant factor $m/2$, eq. (\ref{EnergyHO}) is the usual energy conservation formula for the harmonic oscillator. 

\section{Discussion}

The two examples presented in the previous sections demonstrate a connection between symmetry, here expressed in the form of shape invariance of explicit solutions, and a simple conservation theorem, energy conservation, in a manner that is accessible to all students who are familiar with the basics of Newtonian mechanics. As such, they can serve as a useful tool for introducing beginning students to one of the most elegant results of theoretical physics, and for teaching the gist of Noether's theorem with the help of no more than comparatively basic mathematical tools.

While they provide a broadly accessible illustration, these explicit calculations based on shape invariance do not, of course, provide us with the powerful tool that is Noether's theorem, one of whose key strengths is, after all, that the derivation does {\em not} require explicit solutions. We leave the reader with the following question: The toy models certainly implement the simplified prose version of Noether's theorem, namely that a symmetry leads to a corresponding conserved quantity. In what way, if any, can they be derived directly and rigorously from the standard variational version of Noether's theorem?


\begin{thebibliography}{99}

\bibitem{Noether1918a} Emmy Noether, ``Invarianten beliebiger Differentialausdr{\"u}cke,''  { G{\"o}tt.\ Nachr.} 1918, 37--44.

\bibitem{Noether1918b} Emmy Noether, ``Invariante Variationsprobleme,''  { G{\"o}tt.\ Nachr.} 1918, 235--257.


\bibitem{Byers1996} Nina Byers, ``The Life and Times of Emmy Noether: Contributions of Emmy Noether to Particle Physics,'' in  {\em History of Original Ideas and Basic Discoveries in Particle Physics}, edited by H.B. Newman and T. Ypsilantis (Plenum Press, New York and London, 1996), pp. 945--964.

\bibitem{Kosmann-Schwarzbach2011} Yvette Kosmann-Schwarzbach, {\em The Noether Theorems. Invariance and Conservation Laws in the Twentieth Century}
(Springer, New York, 2011).

\bibitem{Griffiths2008} David Griffiths, {\em Introduction to Elementary Particles} (Wiley-VCH, Weinheim, 2008).

\bibitem{LedermanHill2004}  Leon M. Lederman and Christopher T. Hill, {\em Symmetry and the Beautiful Universe}
(Prometheus, Amherst, 2004).

\bibitem{Wilczek2015} Frank Wilczek, {\em A Beautiful Question: Finding Nature's Deep Design} (Penguin Press, New York, 2015).

\bibitem{DualNote} There are in fact two Noether theorems about the role of symmetry in dynamical systems, which are closely related. The second theorem refers to infinite-dimensional symmetries and how they restrict a system's evolution equation. When physicists talk of ``Noether's theorem'' in the singular case, they usually mean the first theorem, linking symmetries and conservation laws, which is the subject of this article.

\bibitem{Neuenschwander2017} Dwight E. Neuenschwander, {\em Emmy Noether's Wonderful Theorem,} 2nd edition (Johns Hopkins University Press, Baltimore, 2017).

\bibitem{Neuenschwander2014} Dwight E. Neuenschwander, ``Resource Letter NTUC-1: Noether's Theorem in the Undergraduate Curriculum,''
Am. J. Phys. {\bf 82} (3), 183--188 (2014).  DOI: \href{http://dx.doi.org/10.1119/1.4848215}{10.1119/1.4848215}
\end{thebibliography}
\end{document}